\documentclass{article}

\usepackage{graphicx}
\usepackage{amsmath}
\usepackage{flushend}
\usepackage{subfigure}
\usepackage{authblk}

\title{Wideband Antireflection Coating Using Metamaterials}

\author{Fahimeh Sepehripour, Parisa Karimi, and Amin Khavasi}
\affil{Department of Electrical Engineering \\ Sharif University of Technology}

\begin{document}

\maketitle

\begin{abstract}
In this paper, we propose a new approach for realizing antireflection coating using metamaterials. In this approach, a subwavelength array of metallic pillars (with square cross-section) is used for implementing antireflection coating. The effective impedance of the array can be duly adjusted by the size and distance of pillars. Therefore, we design the effective impedance of the antireflection coating to be the geometrical mean of the upper and lower mediums impedance and we choose its height to be a quarter of operating wavelength. Consequently, the reflection vanishes at the desired frequency and fractional bandwidth of 56\% is achieved with a criterion of 10\% reflectance (the refractive index of the substrate is assumed to be 4). The proposed structure is symmetric in both directions. So, it is not sensitive to the polarization of the incident wave at normal incidence. Furthermore, we show that using the multilayer Chebyshev matching transformer of transmission line theory increases the bandwidth of the antireflection up to 107\% at the expense of pass-band ripples. This structure can be used from very low frequencies up to infrared regime by appropriate scaling. 
\end{abstract}

\section{Introduction}
Metamaterials are subwavelength metal-dielectric structures designed to provide extraordinary electromagnetic properties in different frequency ranges. Metamaterials properties are mainly determined by their structures not from the properties of their base compositions \cite{R1,R2}. Metamaterials can be engineered to have the negative refractive index \cite{R3}, the zero refractive index \cite{Rb}, tunable negative permeability \cite{Ra}, high index of refraction \cite{R4}, anomalous dispersion \cite{R5}, optical magnetism \cite{R6} and other electromagnetic properties that cannot be found in the nature. These properties have been widely used in a variety of applications over wide frequency ranges, for instance, THz switches and modulators \cite{R7,R8}, high-resolution magnetic resonance radio-frequency imaging \cite{Rc}, super-directive and super-conductive antennas. Moreover, it has been shown that metamaterials can be tailored to have a desired impedance and this can be used for designing antireflection coatings \cite{Re,Rf,Rg}.

To date, different approaches have been followed for designing antireflection coatings \cite{Rh,Rj,Rk,Rl}. In the classical approach, single or multilayer dielectric films have been used as an antireflection coating \cite{Rm,Rn}. The certain drawback associated with this approach is the difficulty of finding appropriate coating materials that satisfy the index-matching requirement. Metamaterials solve this problem due to their designable effective permittivity. One of the recent demonstrations of a metamaterial antireflection in the infrared regime has been proposed by Kim.et al. in 2015 \cite{R9}. It is a metamaterial structure composed of one-dimensional (1D) metallic slit array. This structure operates as an effective dielectric with designable dielectric permittivity for the polarization perpendicular to the metal grating lines solving the problem of index-matching requirement. However, this antireflection is polarization-sensitive which is not desirable in many applications. Moreover, the bandwidth of this structure is limited to 58\% (with a criterion of 10\% reflectance) for a substrate with refractive index of 4 \cite{R9}.

In this paper, we design a broadband and polarization-independent metamaterial-based antireflection coating using two-dimensional (2D) array of metallic pillars in the infrared regime. Our proposed structure is symmetric in both directions. So, it is not sensitive to the polarization of the incident wave at normal incidence. Furthermore, using the transmission line theory, we show that the structure of multilayer Chebyshev matching transformer increases the bandwidth of the antireflection at the expense of pass-band ripples. Compromising on the flatness of the pass-band response leads to a much more bandwidth performance. Inspired by this concept, we design a two-layer metamaterial-based antireflection coating by using a two-layer Chebyshev impedance transformer, aiming to achieve significant improvement in the bandwidth. We use this approach for both polarization-sensitive structure of \cite{R9} and our polarization-insensitive 2D array of metallic pillars.

This paper is structured as follows: Section \ref{sec2} describes the 1D antireflection coating proposed in \cite{R9} and applying the Chebyshev impedanace transformer appraoch to increase its bandwidth is explained. In Sec.~\ref{sec3}, a polarization-independent antireflection coating is designed. The bandwidth of this structure is increased using the presented  Chebyshev transformer of transmission line theory. The effect of incident angle on the performance of these structures is investigated in Sec.~\ref{sec4}. The paper will be
concluded in Sec.~\ref{sec5}.

\section{1D antireflection coating}
\label{sec2}
\begin{figure}
\centering
\includegraphics[width=2.5in]{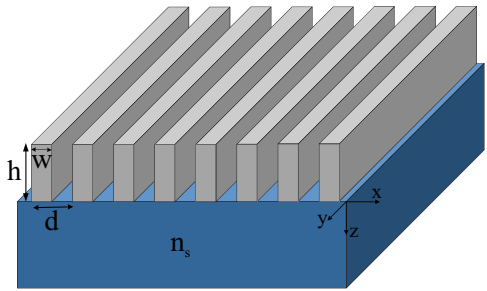}
\caption{Schematic of 1D subwavelength metallic grating metamaterial working as an antireflection coating on a germanium substrate.}
\label{fig1}
\end{figure}

We start this section with the antireflection coating proposed in \cite{R9} which is a metamaterial structure consisting of a one-dimensional metal/air-gap subwavelength silver grating. The substrate, whose refractive index is denoted by $n_s$, is assumed germanium throughout this paper. The schematic of the metamaterial structure is shown in Fig.\ref{fig1} The metallic grating has a subwavelength period of $d$ and the width and the height of the metallic strips are $w$ and $h$, respectively.

\begin{figure}
\centering
\includegraphics[width=8.5cm]{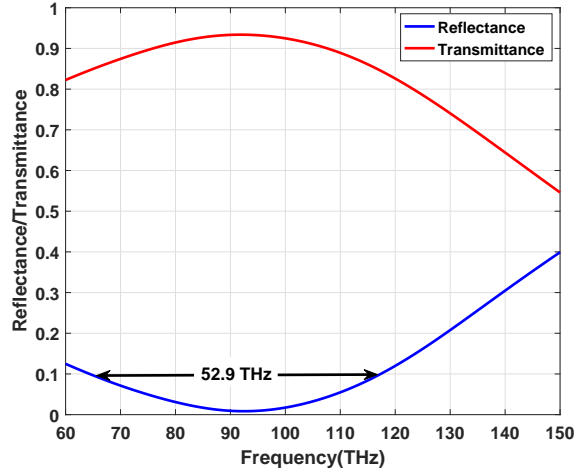}
\caption{Reflectance and transmittance of 1D one-layer grating antireflection versus frequency.}
\label{fig2}
\end{figure}

First, we calculate the transmittance and reflectance of the 1D one-layer silver grating structure on a germanium substrate with the dimensions given in \cite{R9} at $100\,THz$. Here, the transmittance and reflectance of the structure are calculated by finite-difference-time-domain (FDTD) method using a commercial electromagnetic solver (Lumerical FDTD Solutions). The incident wave is polarized in the x-direction while propagating in the z-direction. The optical constant of silver has been extracted from \cite{R10} and the refractive index of the germanium substrate has been assumed to be $n_s=4$. Fig. \ref{fig2} shows the reflectance and transmittance of this structure versus frequency. The bandwidth of this structure (with criterion of reflectance under 10\%) is 58\%.

This structure is equivalent to a transmission line with a quarter-wave transformer $(h = \lambda/4)$ which is a simple and useful circuit for matching a real load impedance to a transmission line. An important feature of the quarter-wave transformer is that it can be extended to multilayer designs in a methodical manner to achieve broader bandwidth by providing smooth transition of characteristic impedances and thus reducing the reflectance. In order to minimize the mismatch, the impedance of each layer must be arranged duly using a multilayer transformer. The most well-known multilayer transformers are Chebyshev and binomial transformers. The Chebyshev multilayer transformer suggests optimum bandwidth at the expense of pass-band ripples and helps to obtain a smooth impedance transition. The Chebyshev transformer is designed by making total reflection coefficient of the transformer $\Gamma(\theta)$ proportional to a Chebyshev polynomial $T_N(sec\theta_mcos\theta)$, where $N$ is the number of sections, $\theta$ is the electrical length of each section and $\theta_m$ is the lower edge of the passband in the transformer. Thus, we have \cite{pozar}:
\begin{equation}\label{eq1}
\begin{split}
\Gamma(\theta) &= 2e^{-jN\theta} \sum_{n=0}^{\lfloor{(N-1)/2}\rfloor} \Gamma_n cos(N-2n)\theta \\ 
&+
\begin{cases}
\frac{1}{2}\Gamma_{N/2} & N\,even \\
0 & N\,odd
\end{cases}\\
&= Ae^{-jN\theta} T_N(sec\theta_m cosn \theta) 
\end{split}
\end{equation}
where $\Gamma_n$ is partial reflection coefficient and $\vert{A}\vert$ is the maximum allowable reflection coefficient magnitude in the pass-band. The value of $\theta_m$  can be calculated as follows \cite{pozar}:
\begin{equation}\label{eq2}
sec\theta_m = cosh\left[\frac{1}{N}cosh^{-1}\left(\frac{1}{\vert{A}\vert}\left\vert\frac{Z_L - Z_0}{Z_L + Z_0}\right\vert\right)\right]
\end{equation}
where $Z_L$ is the load impedance which should be matched to $Z_0$ through the transformer.  For the desired $\vert{A}\vert$ and specified values of $Z_L$ and $Z_0$, $sec\theta_m$ can be determined using Eq.~\ref{eq2} for each selection of $N$. Then, $\Gamma_n$  can be specified from Eq.~\ref{eq1}. Finally, the impedance of other sections can be calculated using \cite{pozar}:
\begin{equation}\label{eq4}
\Gamma_n\simeq\frac{1}{2} ln\frac{Z_{n+1}}{Z_n}
\end{equation}

\begin{figure}
\centering
\includegraphics[width=2.5in]{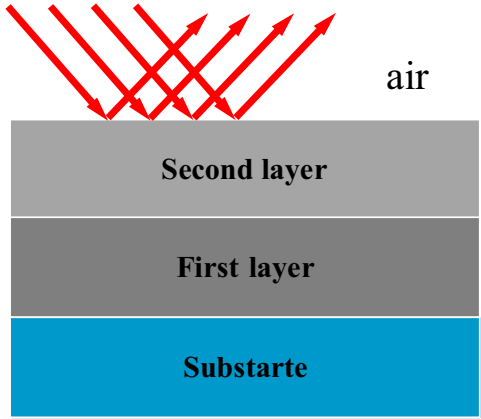}
\caption{Schematic of the two-layer Chebyshev transformer used to design of antireflection coating.}
\label{fig3}
\end{figure}

In this paper, in order to avoid increasing the size of the structure and fabrication difficulties we choose a two-layer structure ($N=2$) to improve the bandwidth of the antireflection coating as shown in Fig.~\ref{fig3}. Now, $Z_L$ and $Z_0$ are the free space and the substrate impedances given by:
\begin{equation}\label{eq3}
\begin{split}
Z_L &= \eta_0 \\
Z_0 &= \frac{\eta_0}{n_s}\\
\end{split}
\end{equation}
where $\eta_0$ is the characteristic impedance of the free space. The substrate is germanium with $n_s  = 4$ and $A = \sqrt{0.1}$ is assumed to have reflected power under 10\%. It should be noted that $\theta$ is equal to $\pi/2$ in the Chebyshev transformer which means that the height of each layer is equal to a quarter wavelength. The calculated impedances of each layer normalized to the free-space impedance are presented in Table.~\ref{table1}.

\begin{table}[!t]
\renewcommand{\arraystretch}{1.5}
\caption{Calculated normalized impedances of each layer in a two-layer antireflection coating}
\label{table1}
\centering
\begin{tabular}{|l||*{5}{c|}}
\hline
two-layer antireflection & First-layer & Second-layer\\
\hline
Normalized-impedance & 0.4141 & 0.6037\\
\hline
\end{tabular}
\end{table}

\begin{figure}
\centering
\includegraphics[width=2.5in]{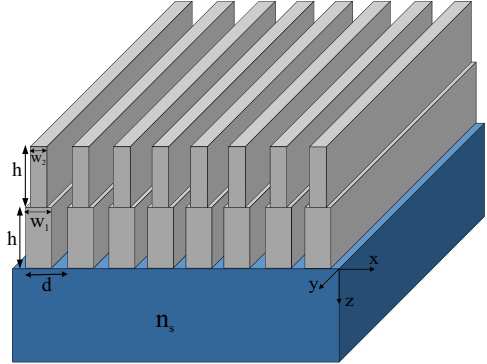}
\caption{Schematic of 1D two-layer metallic grating on a germanium substrate. The height of both layers are equal but the width of metals in each layer is chosen in accordance with the desired impedance.}
\label{fig4}
\end{figure}

The first and second layer of the structure shown in Fig.~\ref{fig3} can be implemented by subwavelength one-dimensional metallic gratings as shown in Fig.~\ref{fig4}. These gratings have identical periods but the width of metals in each layer is determined based on the desirable impedance given in Table.~\ref{table1}. It is worth mentioning that the impedance of each layer (for x-polarized incident wave) in 1D subwavelength metallic grating is related to its $w_i/d$ ratio by the following relation \cite{Ro}:
\begin{equation}\label{eq9}
Z_i = \eta_0(1-\frac{w_i}{d})\,\,\,\,\,\,\,i=1,2
\end{equation}
where $d$ is the period of the gratings and $w_i$ is the width of the metals in each layer specified by $i=1,2$. In Eq.~\ref{eq9}, it is assumed that the medium filling the slits is air. Given the impedance of each layer as listed in Table. \ref{table1}, the dimensions of metallic strips are specified by Eq.~\ref{eq9} as follows: $w_1=0.6037d$, $w_2=0.4141d$ and $h_1=h_2$ are equal to quarter of the wavelength. Let us assume that the center frequency for our design is $100\,THz$. We set the subwavelength period of the metallic gratings to $d=\lambda/10=300\,nm$.Therefore, The height of both layers will be $750nm$.

\begin{figure}
   \centering
    \subfigure[]
    {
        \includegraphics[width=8.5cm]{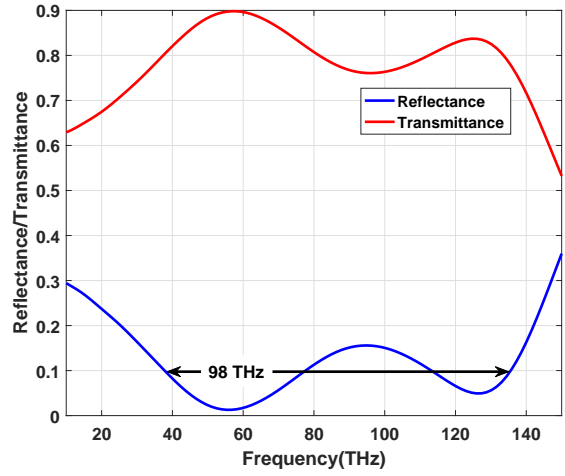}
        \label{fig5-1}
        \hfil
    }
    \\
    \subfigure[]
    {
        \includegraphics[width=8.5cm]{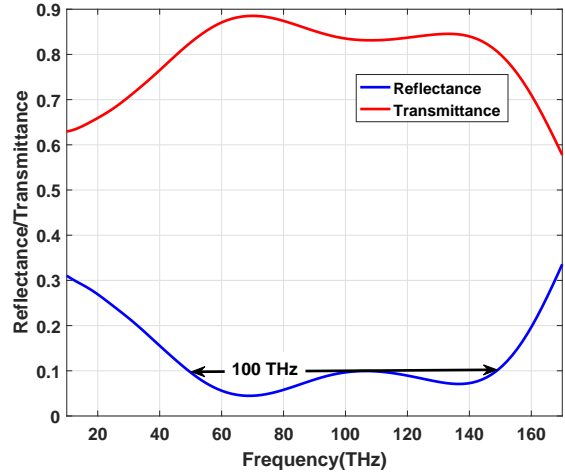}
        \label{fig5-2}
        \hfil
    }
   \centering
  \caption{Reflectance and transmittance of the structure shown in Fig.~\ref{fig4} versus frequency with (a) initial dimensions (b) slightly changed dimensions.}
  \label{fig5}
\end{figure}

Reflectance and transmittance for an x-polarized incident wave are plotted in Fig.~\ref{fig5-1}. It can be seen that the reflectance of the two-layer structure has a peak higher than $10\%$ in the desired bandwidth and this is due to neglecting the effect of the higher diffraction orders in the Eq.~\ref{eq9}. This issue can be resolved by fine-tuning the dimensions of the structure. Another problem is matching the obtained center frequency the desired center frequency ($100\,THz$). In order to shift the center frequency of the reflectance and control the reflectance level, the height and width of the pillars should be changed, respectively. So, the reflectance and transmittance of the redesigned two-layer structure are shown in Fig.~\ref{fig5-2} for the following dimensions which slightly changed:  $h = 650\,nm$, $d=260\,nm$, $w_1=145\,nm$,and $w_2=80\,nm$. In this figure the central frequency is $100\,THz$, and the reflectance is totally lower than $10\%$ in the bandwidth.

It is clear that the bandwidth of the two-layer structure is substantially greater than the one-layer one. Although the bandwidth of two-layer structure is increased up to $100\%$, this structure is polarization sensitive. In the next section, we present a simple solution to this problem. 

\section{Antireflection coating using an array of metallic pillars }
\label{sec3}
\begin{figure}
\centering
\includegraphics[width=2.5in]{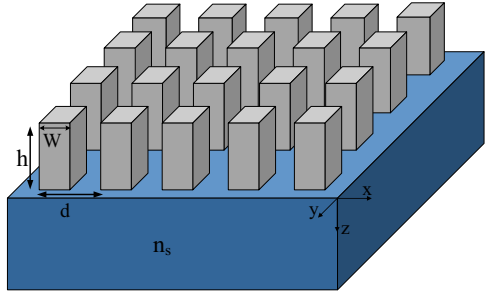}
\caption{Schematic of one-layer subwavelength array of square pillar array on a substrate designed to operate as a polarization-independent antireflection coating.}
\label{fig6}
\end{figure}

The schematic of the proposed polarization-independent antireflection coating is shown in Fig.~\ref{fig6}. This periodic structure is made of metallic pillars with square cross-section on a substrate with refractive index $n_s$. The two-dimensional (2D) square pillar array has a subwavelength period of $d$ in both directions. The side length and height of the pillars are $w$ and $h$, respectively. The square pillar array is arranged in a square lattice and the structure is symmetric. As a result, this structure is polarization-independent at normal incidence.

The array of metallic pillars can be modeled by an uniaxial anisotropic effective medium by attributing a surface conductivity at the interface of the structure with the surrounding media. The surface conductivity is due to higher diffracted orders and can be ignored when the period of the structure is much smaller than the wavelength \cite{R11}. Accordingly, the effective electric permittivity and permeability of the antireflection coating can be obtained from the following equations \cite{R11}:
\begin{equation}\label{eq11}
\begin{aligned}
\mu_x = \mu_y = S_0^2\\
\epsilon_x = \epsilon_y = \frac{1}{S_0^2}
\end{aligned}
\end{equation}
where $S_0$  is the overlap integral between the zeroth-order plane wave and the component of the electric field of the TEM mode that is parallel to the incident electric field $p(x.y)$ across the whole unit cell and it is calculated as follows \cite{R11}:
\begin{equation}\label{eq12}
S_0^2= \frac{[\iint_{unit\,cell}{p(x,y)dxdy}]^2}{d^2\iint_{unit\,cell}{p^2(x,y)dxdy}}
\end{equation}
An analytical approximate expression for $p(x,y)$ has been given in \cite{R11} when the pillars are perfect electric conductors (PECs).

According to Eq.\ref{eq11} the effective impedance of the structure is equal to $Z_{eff}= \eta_0 S_0^2$, where $\eta_0$ is characteristic impedance of the free space. It is worth mentioning that $S_0^2$  is only dependent on the $w/d$ ratio. Consequently, by appropriate designing of the $w/d$ ratio, the required impedance of the antireflection coating can be acquired. Values of normalized impedances for different $w/d$ have been presented in Table.~\ref{table2}.

\begin{table*}
  \centering
  \caption{Calculated $w/d$ ratios to obtain various effective characteristic impedance for square pillar array normalized to the characteristic impedance of free space}
\renewcommand{\arraystretch}{1}
\begin{tabular}{|l||c|c|c|c|c|c|c|c|c|}
\hline
Normalized Impedance & 0.1 & 0.2 & 0.3	& 0.4 & 0.5 & 0.6 & 0.7 & 0.8 & 0.9 \\
\hline
$w/d$ & 0.8& 0.78&	0.73&0.66&0.56&0.45&0.34&0.22& 0.104 \\
\hline
  \end{tabular}
  \label{table2}
\end{table*}

For a one-layer antireflecion coating the impedance of the coating should be the geometric mean of the characteristic impedance of the substrate and free space. Hence, we need $Z_{eff}=\eta_0/\sqrt{n_s}$. Assuming $n_s=4$ and according to Table.~\ref{table2} we obtain $w=0.56d$. As already pointed out, the height of antireflection coating is set to $h=\lambda/4$. It should be noted that the square pillars should have a sufficiently subwavelength period. Thus, the period of the structure is fixed at $d=\lambda/10$.

\begin{figure}
\centering
\includegraphics[width=8.5cm]{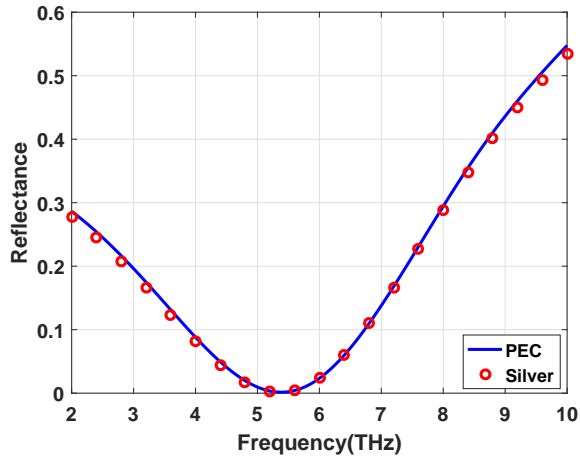}
\caption{Reflectance of the one-layer antireflection coating versus frequency assuming metals as PEC (blue solid line) and silver (red circles).}
\label{fig7}
\end{figure}

First, we consider the design of antireflection coating at low frequencies where the metals can be approximated by PEC. Our simulations show that PEC assumption is accurate up to the terahertz regime. This is demonstrated in Fig.~\ref{fig7} where the reflectance of the antireflection coating designed at frequency of $5\,THz$ is plotted by assuming metals to be PEC (blue solid line) and silver (red circles).

The results of Fig.~\ref{fig7} show that zero reflectance is observed around the designed center frequency $5\,THz$. In this case, the bandwidth is $54\%$. The slight deviation in the minimum reflectance frequency from what we expect from the circuit model is due to neglecting the effect of higher diffraction orders. This negligible discrepancy can be easily resolved by $3\%$ increase in the height of the pillars. 

\begin{figure}
\centering
\includegraphics[width=8.5cm]{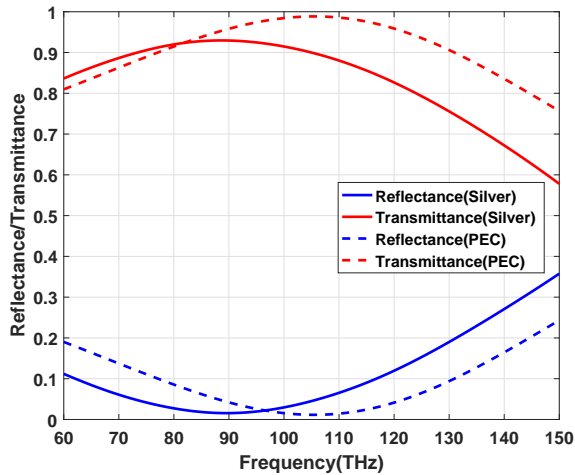}
\caption{Reflectance and transmittance of one-layer antireflection coating versus frequency in the infrared regime assuming metals as PEC (dashed lines) and silver (solid lines).}
\label{fig8}
\end{figure}

At infrared frequencies, PEC approximation is not very accurate. As an example, consider an antireflection coating designed at $100\,THz\,(\lambda= 3\mu m)$ with $d=\lambda/10=300 nm$, $w=0.56d=168nm$ and $h=\lambda/4=750nm$. Fig. \ref{fig8} shows the transmittance and reflectance of this structure versus frequency when the pillars are assumed as silver (solid lines) and PEC (dashed lines). It can be seen that when the metals are assumed as silver, the minimum reflectance appears at $90\,THz$ which is fairly close to the desired central frequency $100\,THz$. On the other hand, the normalized bandwidth of this structure is slightly increased to $56\%$ due to the loss of the silver which slightly reduces the amount of reflectance (and also transmittance).

\begin{figure}[t]
\centering
\includegraphics[width=8.5cm]{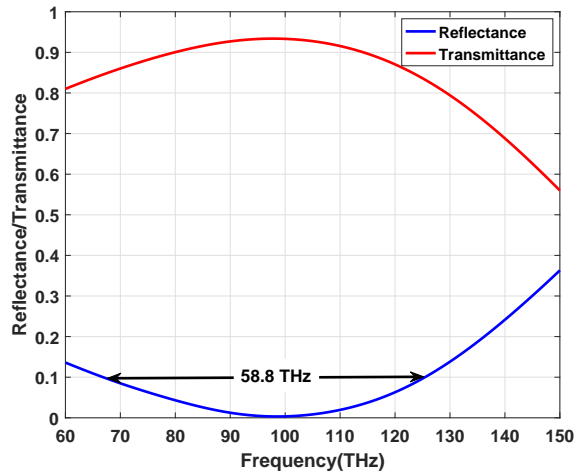}
\caption{Reflectance and transmittance of optimized one-layer square pillars versus frequency in the infrared region.}
\label{fig9}
\end{figure}

The difference between the desired center frequency and the minimum level of reflectance can be resolved with a slight adjustment of structure dimensions. By reducing the height of the pillars $(h= 700\,nm)$ and increasing the width of the pillars $(w = 180\,nm)$, the minimum reflectance frequency is shifted to $100\,THz$. Also, the minimum reflectance is almost zero. The simulation results of this structure have been shown in Fig.~\ref{fig9}. In this case, the bandwidth of almost $59\%$ is obtained. 

\begin{figure}[t]
\centering
\includegraphics[width=2.5in]{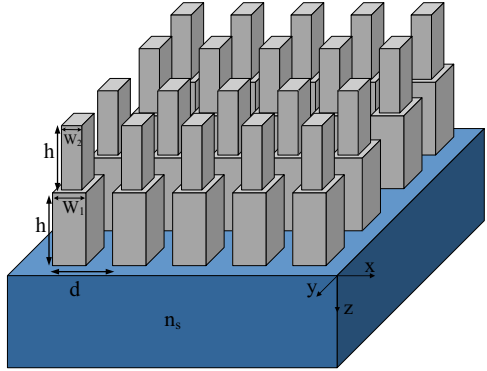}
\caption{Schematic of 2D two-layer subwavelength metamaterial structure acting as an antireflection coating.}
\label{fig10}
\end{figure}

Now, as shown in Fig.~\ref{fig10}, we design a two-layer antireflection coating based on Chebyshev transformer to increase the bandwidth. The impedance of each layer is given in Table. \ref{table1}. According to Table. \ref{table2}, these impedances can be realized by $w_1  = 0.63d$ and $w_2  = 0.46d$ where $d = \lambda/10$ is chosen. Furthermore, the height of each layer for a Chebyshev transformer is $h = \lambda/4$.

\begin{figure}[t]
   \centering
    \subfigure[]
    {
        \includegraphics[width=8.5cm]{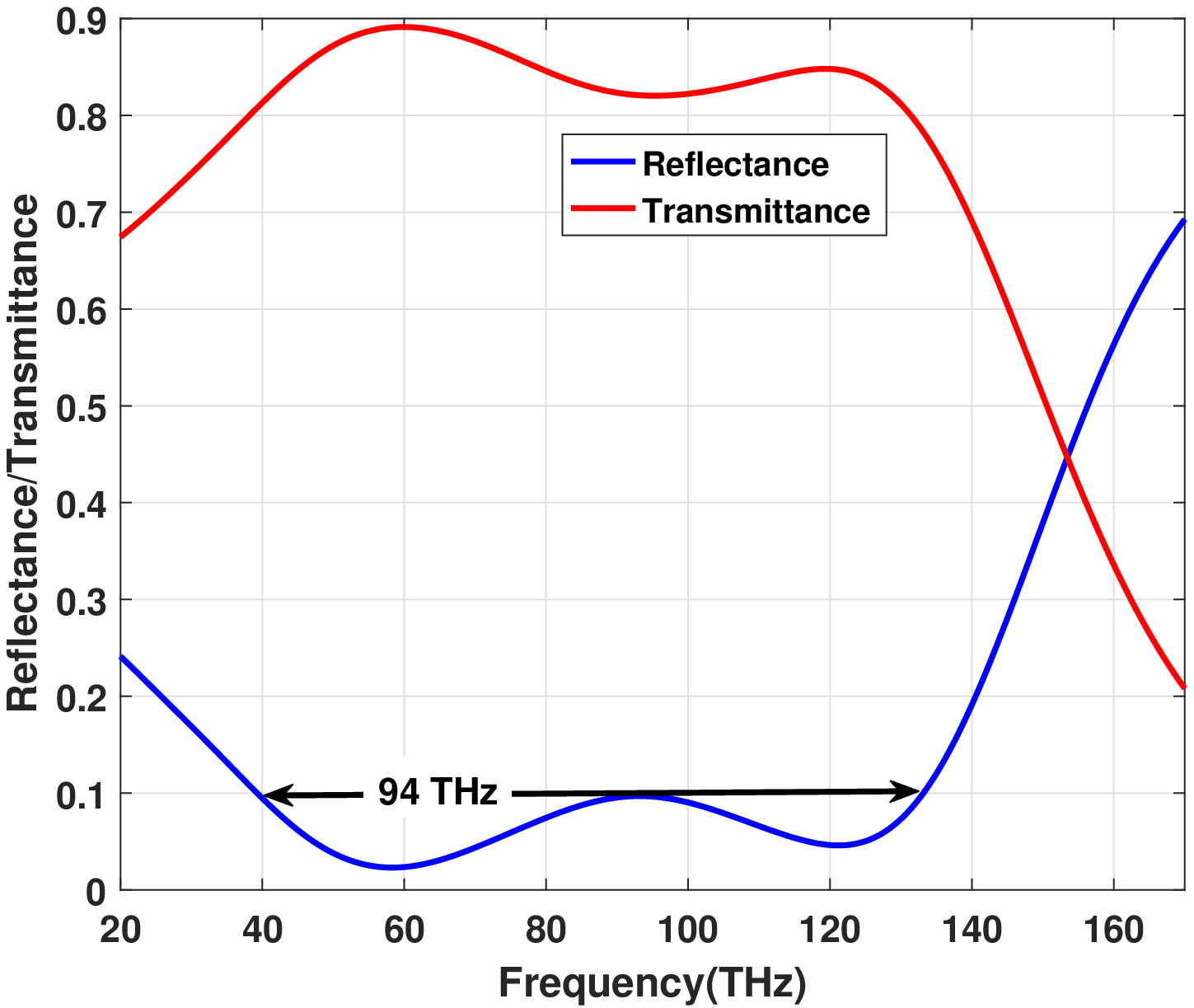}
        \label{fig11-1}
        \hfil
    }
    \\
    \subfigure[]
    {
        \includegraphics[width=8.5cm]{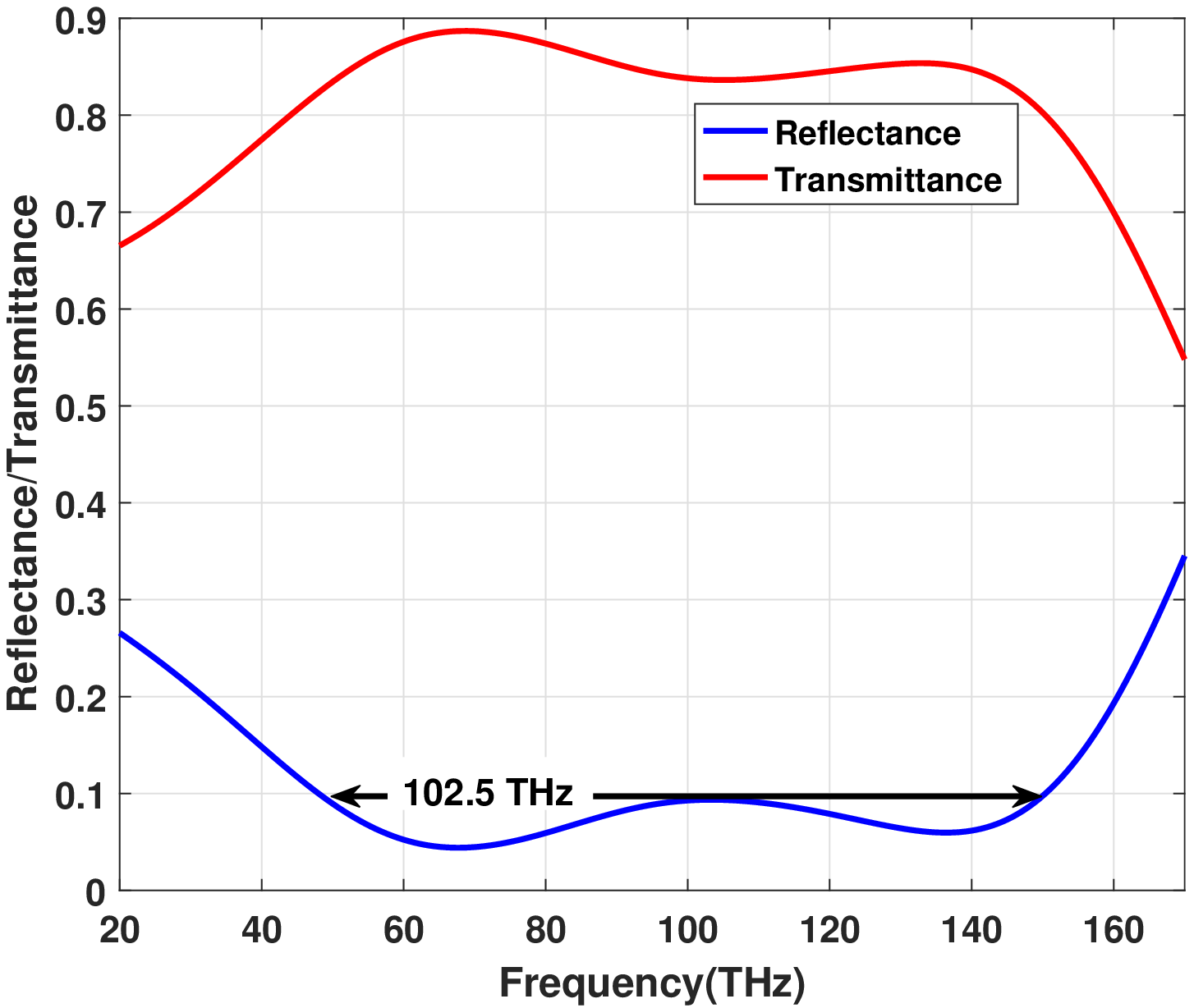}
        \label{fig11-2}
        \hfil
    }
   \centering
   
  \caption{The Reflectance and the transmittance of two-layer 2D antireflection coating versus frequency (a) before fine-tuning (b) after fine-tuning.}
  \label{fig11}
\end{figure}

Reflectance and transmittance of two-layer structure are plotted in Fig.~\ref{fig11-1}. Same as the 1D structure the center frequency of the result do not match to the desired center frequency $100\,THz$. In order to increase the center frequency of the reflectance curve, the height of the pillars should be reduced. The dimensions of the two-layer structure are tuned as $h = \lambda/4 = 640\,nm$, $w_1= 0.63d$ and $w_2  = 0.46d$ with $d =\lambda/10$. Using these fine-tuned dimensions, the structure shown in Fig.~\ref{fig10} is simulated and the transmittance and reflectance of this structure versus frequency are shown in Fig.~\ref{fig11-2}. In this figure, the pillars are assumed silver and the maximum bandwidth is equal to $103\%$. In the next section, we investigate the sensitivity of this structure to the angle of the incident wave. 

\section{ Oblique incidence}	
\label{sec4}
\begin{figure}[t]
   \centering
    \subfigure[]
    {
        \includegraphics[width=8.5cm]{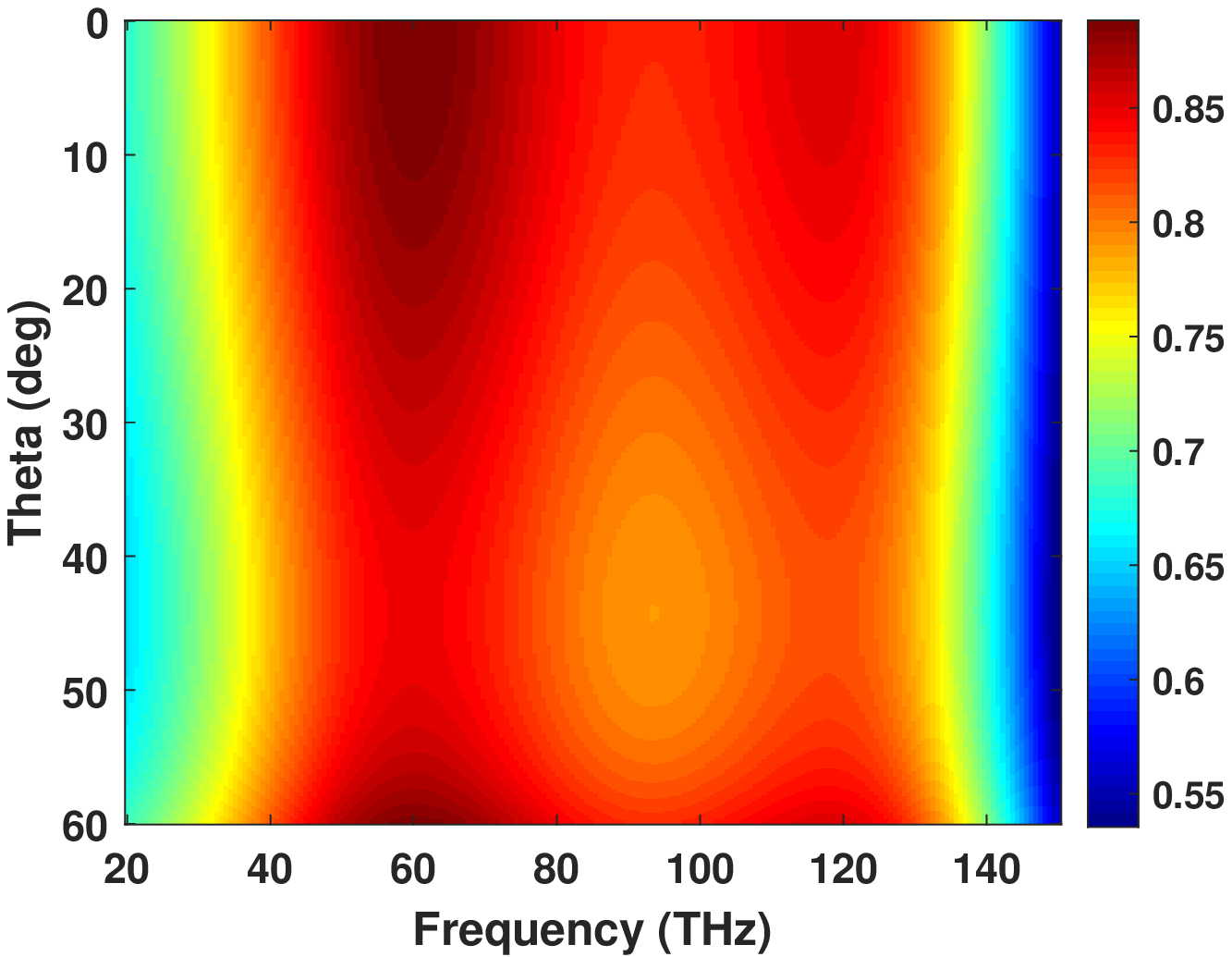}
        \label{fig13-1}
        \hfil
    }
    \\
    \subfigure[]
    {
        \includegraphics[width=8.5cm]{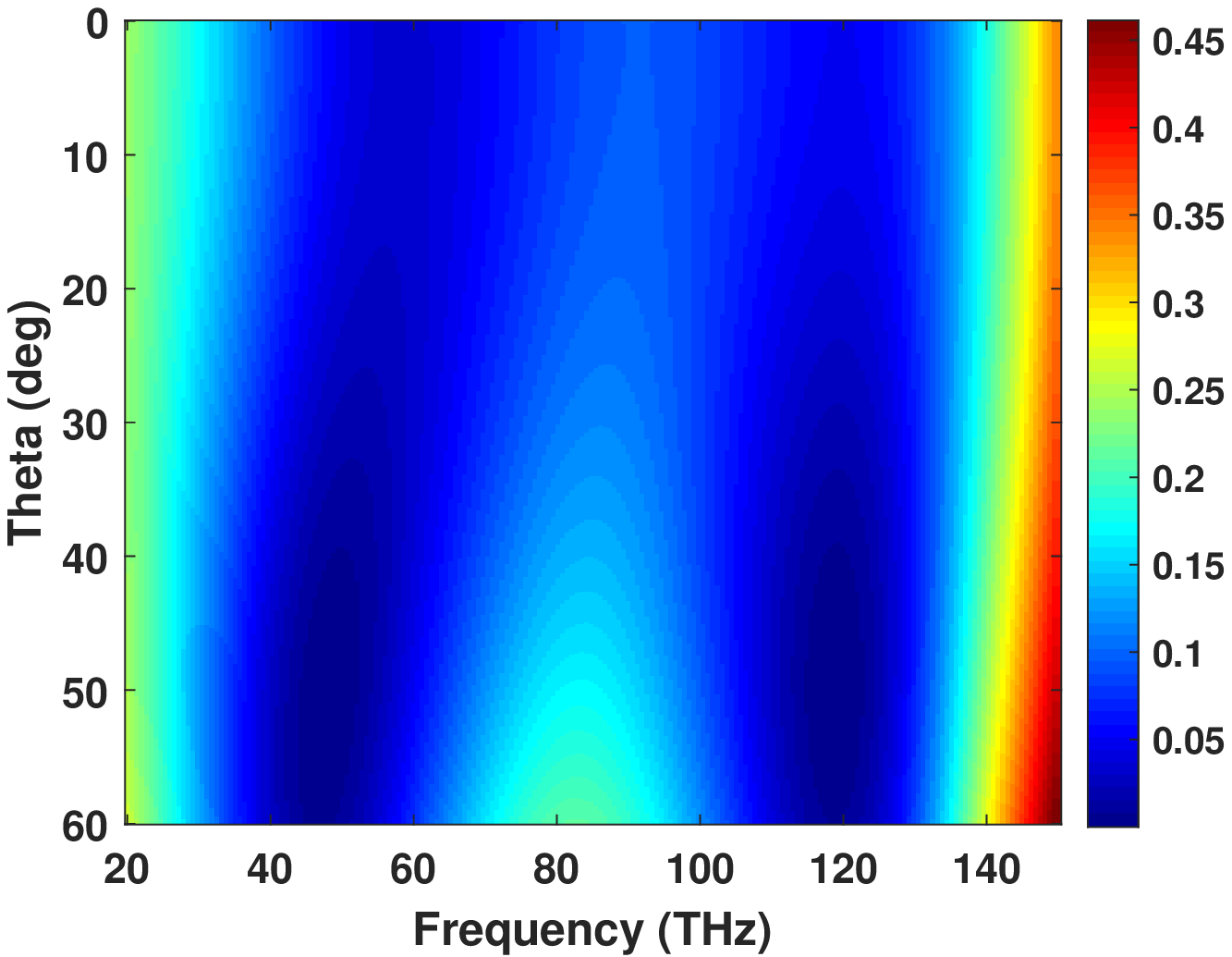}
        \label{fig13-2}
        \hfil
    }
  \centering
  \caption{(a) Transmittance and (b) Reflectance of the proposed 2D two-layer antireflection coating versus frequency and angle of incidence for TE polarization}
  \label{fig12}
\end{figure}

\begin{figure}[t]
   \centering
    \subfigure[]
    {
        \includegraphics[width=8.5cm]{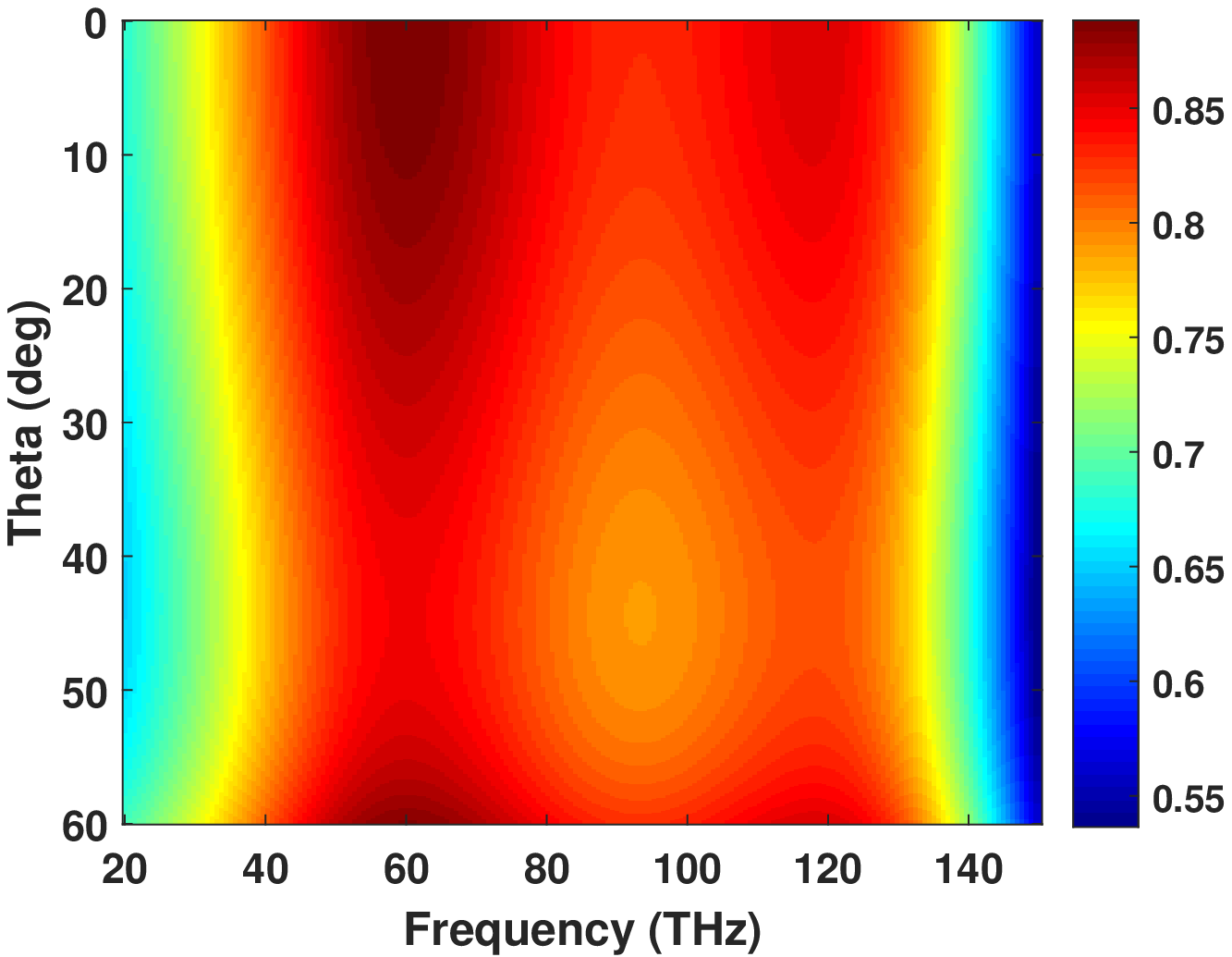}
        \label{fig13-3}
        \hfil
    }
    \\
    \subfigure[]
    {
        \includegraphics[width=8.5cm]{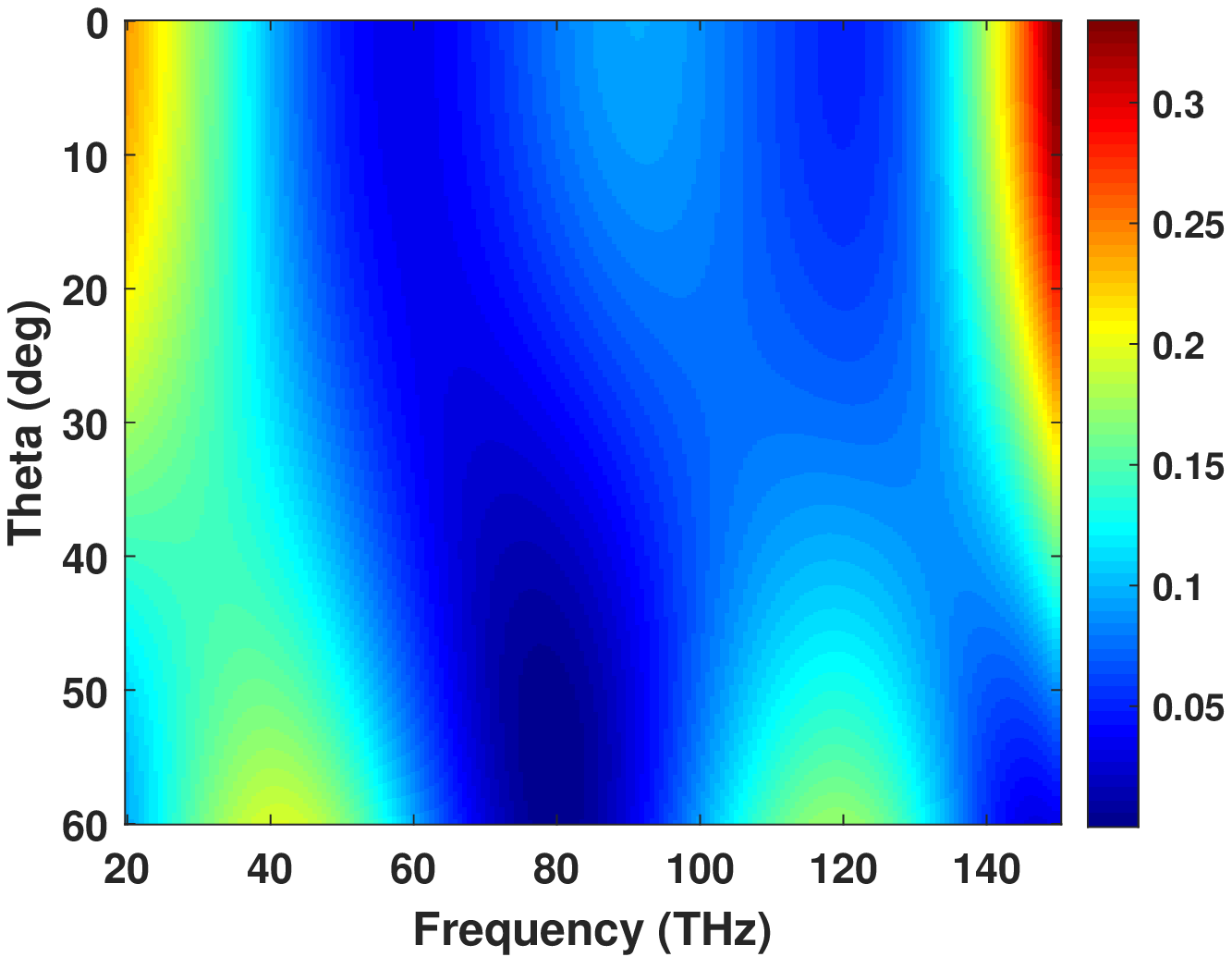}
        \label{fig13-4}
        \hfil
    }
  \centering
  \caption{(a) Transmittance and (b) Reflectance of the proposed 2D two-layer antireflection coating versus frequency and angle of incidence for TM polarization}
  \label{fig13}
\end{figure}

We also study the dependence of the metamaterial-based antireflection coating performance to the angle of the incident wave for the structure shown in Fig.~\ref{fig10} for both TE and TM-polarizations. Fig.~\ref{fig13-1} and Fig.~\ref{fig13-2} show the transmittance and reflectance versus frequency and angle of incidence in the y-z plane for TE polarization. On the other hand, Fig.~\ref{fig13-3} and Fig.~\ref{fig13-4} show the transmittance and reflectance versus frequency and angle of incidence in the y-z plane for TM polarization. It is found that the performance of the proposed antireflection remains good within the incident angle range from $0$ to $30$ degree for TE polarization and from $0$ to $40$ degree for TM polarization.

\section{Conclusion}
\label{sec5}
In this paper, we have presented that a novel broadband metamaterial antireflection coating can be designed by using the multilayer Chebyshev matching transformer in wide frequency ranges. Firstly, we increased the bandwidth of the 1D metamaterial-based antireflection coating which has been proposed by Kim et al. up to 100\% in the infrared regime using this idea. The 1D structure is polarization sensitive. For implementing a polarization-sensitive antireflection coating, we present a metamaterial-based structure which consists of a sub-wavelength array of metallic pillars with a square cross-section. The dimensions of this structure were specified by utilizing analytical relations. So the effective impedance of the metallic pillar array is the geometric mean of the air and the substrate impedance to eliminate the reflectance at the desired frequency. The thickness of the subwavelength metallic pillars is determined a quarter of the free space wavelength. The fractional bandwidth of the proposed one-layer antireflection coating for germanium substrate is equal to 56\%. Furthermore, by using the two-layer Chebyshev matching transformer, the bandwidth of the structure is increased to 107\%. At normal incidence, this structure is polarization independent due to the symmetry of the structure. Also, we have investigated the sensitivity of the antireflection coating performance to the angle of incidence and acceptable performance at oblique incidence was observed.

\bibliographystyle{unsrt}
\bibliography{References}

\end{document}